\documentclass[nohyper,12pt,letterpaper]{JHEP}
\usepackage{epsfig}

\newfont{\frak}{eufm10 scaled 1200}

\newfont{\Bbb}{msbm10 scaled 1200}     
\newcommand{\mathbb}[1]{\mbox{\Bbb #1}}
\DeclareSymbolFont{AMSa}{U}{msa}{m}{n}
\DeclareSymbolFont{AMSb}{U}{msb}{m}{n}
\let\Box\relax
\DeclareMathSymbol{\Box}{\mathord}{AMSa}{"03}

\def \t{\theta}
\def \b{\beta}

\def \f{\phi}

\def \p{\pi}

\title{The Interplay between $\theta$ and $T$}

\author{W. Fischler, E. Gorbatov, A. Kashani-Poor, R. McNees, S.
Paban, P. Pouliot\\
  Department of Physics\\
  University of Texas, Austin, TX 78712\\
E-mail: \email{fischler, elie, kashani, mcnees \\
\hspace{1.2cm}paban, pouliot@physics.utexas.edu}}

\abstract{We extend a recent computation of the dependence of the free
energy, F,
 on the
noncommutative scale $\theta$ to theories with very different UV
sensitivity.  The temperature dependence of $F$ strongly
suggests that a reduced number of degrees of freedom contributes to the free
energy in the non-planar sector, $F_{\rm np}$, at high temperature.

This phenomenon seems generic, independent of the UV sensitivity, and can be
traced to modes whose thermal wavelengths
become smaller than the noncommutativity scale. The temperature dependence of
$F_{\rm np}$ can then be calculated at high
temperature using
classical statistical mechanics, without encountering a UV catastrophe even
in large
number of dimensions. This result is a telltale sign of the low number of
degrees
of freedom contributing to $F$ in the non-planar sector at high
temperature. Such
behavior is in marked contrast to what would
happen in a field theory with a random set of higher derivative
interactions.}


\preprint{\hepth{0003216}\\ UTTG-05-00 \\}
\begin{document}


\section{Introduction}
Noncommutative field theories have proven to exhibit fascinating properties
(\cite{Gopakumar:2000zd}-\cite{Filk:1996dm}).

As we noted in a recent paper \cite{Fischler:2000fv}, there is a drastic
reduction in the number of degrees of freedom that contribute to the free
energy
in the non-planar sector, $F_{\rm np}$, at high temperatures. By the free
energy
in the non-planar sector, we mean those contributions to the free energy which
arise from non-planar Feynman graphs. This reduction can be read off
by looking at the temperature dependence of $F_{\rm np}$.
In this paper, we will present further evidence for this reduction, and
study two further theories with very different UV sensitivity to test how
generic
this high temperature behavior is.

We will take $\t_{0i}=0$ throughout.

In \cite{Fischler:2000fv}, we also noted the existence of winding states in
noncommutative field theories. Though we have no clean way of treating
these modes
separately from the conventional degrees of freedom of commutative field
theory,
it appears plausible that $F_{\rm np}$ is indicative of
these additional degrees of freedom, especially at high temperature, at which
the winding states become light.

In studying the reduction of the degrees of freedom that contribute to $F_{\rm
np}$, we emphasize that this phenomenon does not depend on the sensitivity
of the
corresponding commutative theory to the ultraviolet \cite{Matusis:2000jf}.
Indeed,
this reduction occurs in theories that are extremely insensitive to the
ultraviolet, like the $ N = 4$, $D = 4$ SYM theory, as well as theories with
strong ultraviolet sensitivity, as eg. $\lambda
\phi^3$ in $D=6$.

In a nutshell, what we think happens is that as the temperature rises above the
noncommutativity scale, the modes of the various fields with momenta of
order the temperature, $T$, do not contribute to $F_{\rm np}$. This occurs
because, as the thermal wavelengths drop below the noncommutativity scale
$\t^{1/2}$, as far as their contribution to $F_{\rm np}$ is
concerned, there is no way to distinguish these high momenta modes from each
other and thus no way to count their contribution individually. Another way of
thinking about this is that the contributions to $F_{\rm np}$ are sensitive to
the phase space structure of space: the uncertainty principle
among the spatial coordinates
$x_i$ renders the
identification of the degrees of freedom with regard to their contribution to
$F_{\rm np}$ impossible; this contribution seems to be summarized by a few
quantum degrees of freedom contributing per noncommutative cell whose area
is $\t
$. In the following, we will refer to this cell as the Moyal cell.

In the next section, we briefly review the results for the free energy of the
Wess-Zumino model. We introduce a tool to test whether a reduction of
degrees of
freedom contributing to $F_{\rm np}$ in fact takes place,
the calculation of the classical statistical mechanics approximation to F, and
discuss when we expect this approximation to be valid.

In the third section, we discuss the case of $N=4$, $D=4$ SYM. This theory
also shows a
reduction of degrees of freedom at high temperatures, although the theory
is rather
insensitive to the UV. We see that the nonplanar sector again contains winding
states.

In the fourth section, we show how these results generalize to higher
dimensions
where there can be more noncommutative spatial directions. We show that the
classical approximation remains valid for $\lambda \phi^3$ in spite of the high
sensitivity of the corresponding commutative theory to the UV.

Finally, we end with conclusions.

\section{The thermodynamics of the noncommutative Wess-Zumino model in
perturbation
theory}

The Lagrangian density for the Wess-Zumino model is:

$$ {\cal L} = i \partial_\mu \bar{\psi}\bar{\sigma}^\mu \psi + A^* \Box\, A
-\frac{1}{2} M
\psi
\psi -\frac{1}{2}M \bar{\psi}\bar{\psi} -g \psi * \psi\,  A -
g\bar{\psi}*\bar{\psi}\, A^* -F^*  F \,,$$
where $F$ is given by
$$F= - M A^* - g A^* * A^* \,.$$

We showed in a previous paper that in the noncommutative Wess-Zumino
model, one can
distinguish two regimes of high temperature, $\b M \ll 1$.\footnote{We will
assume that the Compton wavelength of the fields is bigger than the
noncommutative scale, $\t M^2
\ll1$.} These two regimes are: $\t T^2 \ll 1$ and $ \t T^2 \gg 1$.

In the case $\t
T^2 \ll 1$,
the thermal wavelength is larger than the noncommutativity scale. Here, $F_{\rm
np}$ behaves as

\begin{equation}
\frac{F}{V} {\Big|_{\rm np}} \sim - g^2 T^4 \,.
\end{equation}
In the other limit, the free energy density behaves as

\begin{equation}
\frac{F}{V}  {\Big|_{\rm np}}  \sim  -g^2 \frac{T^2}{\t} \log{T^2 \t} \,.
\label{red}
\end{equation}
This expression for the free energy shows a dramatic reduction of the
degrees of freedom contributing to $F_{\rm np}$.

Such a reduction has a natural interpretation in terms of the novel phase space
structure due to the noncommutative nature of space. At any temperature, the
typical momenta of the fields have magnitudes of order the temperature or
smaller. As the temperature rises, the thermal wavelengths become smaller. When
the temperature reaches values of $O(\t^{-1/2} )$, modes of the field with
momenta
of
$O(T)$ no longer contribute to $F_{\rm np}$.

More precisely, what we think happens is that at temperatures $T^2 \gg1/\t$,
the modes
$\f_{k_1,k_2,k_3}$ with $ k_{1,2} \geq \t^{-1/2} $ cannot be distinguished
from each
other insofar as their contribution to $F_{\rm np}$ is concerned, and therefore
will not be counted individually in the non-planar contribution to the
partition
function. These modes lose their individual identity in the non-planar
sector because the noncommutativity between
$x_1$ and $x_2$, $[x_1,x_2] \neq 0$, implies an uncertainty relationship which
renders impossible their separate identification. This
may explain the reduction in the temperature dependence of the free energy from
$T^4$ to $T^2/\t$.

The non-planar sector yields a contribution to the free energy which, could
this sector be isolated, would suggest that it has the number of degrees of
freedom of a 1+1 dimensional field theory at temperatures
$T \gg1/\t^{1/2}$. This is reminiscent of the result obtained by Atick and
Witten
\cite{Atick:1988si} in the context of string theory. There is also a
subleading,
logarithmic dependence on the temperature in
equation (\ref{red}) for the
free energy, which is all that remains from the contributions of the high
momenta components
of the fields.

As announced in the introduction, one can compare this computation to a
classical statistical mechanics calculation  of the free energy. The
classical calculation can be done by just keeping the zero frequency term
in the sum over frequencies. The result is

\begin{equation}
\frac{F}{V}  {\Big|_{\rm np}} \sim - g^2 \frac{T^2}{\t}\log \Lambda \,,
\end{equation}
where $\Lambda$ is an ultraviolet cutoff.

The classical calculation captures the correct
power law dependence in the temperature
but replaces the logarithmic dependence on temperature obtained in quantum
mechanics by $\log \Lambda$.

We should remind the reader that classical statistical mechanics is a good
approximation at high temperatures for the free energy of systems with few
degrees of freedom. For example, the high temperature behavior of the free
energy
for a particle of mass
$m$ in a potential
$V(x)$ is  well approximated by classical statistical mechanics when the
thermal
 de-Broglie wavelength ${\lambda}_{DB} \sim \frac{1}{(mT)^{1/2}} $ is smaller
than the length scale over which the potential varies. In the path integral
formulation, this would correspond to actually shrinking the temporal
circle down
to zero size and performing a dimensional reduction. In Feynman diagrams, this
amounts to only keeping the contributions of the zero frequency components
in the
sum over frequencies.

In contrast, if one uses classical statistical mechanics to evaluate the free
energy of systems with a field theory number of degrees of freedom at high
temperature, one inevitably encounters a UV catastrophe.

On the other hand, classical statistical mechanics is a good approximation in
field theory when calculating thermal correlation functions. An example is the
calculation of the two point correlation function between two operators,
$O_1(x)$
and $O_2(y)$,

$$ \langle O_1(x)O_2(y) \rangle - \langle O_1(x) \rangle \langle O_2(y) \rangle
\;.$$ Indeed, the modes that primarily contribute to this correlation function
have wavelength commensurate with the distance $|x - y|$ separating the
probes. As one heats the system, the population of these modes increases
and hence
their contribution to the correlation function is well approximated by
classical
statistical mechanics. So the classical approximation at high temperature
becomes
applicable in field theory when there is a bound on the wavelengths of the
modes
that contribute to the quantity we are calculating.

The fact that $F_{\rm np}$ is well approximated by
classical statistical mechanics thus suggests that only a reduced set of
degrees
of freedom contributes to $F_{\rm np}$ at high temperatures. This should be
contrasted to the cases of field theories with an arbitrary infinite series of
higher derivative terms in the Lagrangian. In such generic cases, if one
attempts
to approximate the free energy by a classical statistical calculation, one
encounters enhanced UV catastrophes as compared to the theory without the
higher
derivative terms. The averting of the UV catastrophe due to the special
choice of
higher derivative terms in noncommutative field theories is therefore quite
remarkable.

\section{The thermodynamics of $N=4$, $D=4$ Super Yang-Mills}

Our interest in examining this case is to see how crucial the ultraviolet
sensitivity of a
noncommutative theory is to the existence of winding states and the high
temperature behavior of $F_{\rm np}$. We will show that the existence
of winding states as well as the reduction at high temperatures in the
number of degrees of
freedom contributing to $F_{\rm np}$ are independent of
the UV behaviour of the noncommutative field theory.

The calculation of the free energy is straightforward, the result can be
found in the
literature \cite{Fotopoulos:1998es,Arcioni:2000hw}. We present for completeness
some of the steps in the calculation of the free energy in appendix \ref{appa}.

The contribution to the free energy coming from the nonplanar sector is:

\begin{equation}
\frac{F}{V}  {\Big|_{\rm np}}= -4 g^2 N \int \frac{d^3 p}{(2\p)^3} \frac{d^3
k}{(2\p)^3}
\frac{e^{i\, p\, \t\, k}
}{4
\, \omega_p
\, \omega_k} \Big(n_B (\omega_p) + n_F(\omega_p)\Big)\, \Big(n_B(\omega_k) +
n_F(\omega_k)\Big)   \label{sym}.
\end{equation}
Up to an overall factor, this is exactly the result we found in the Wess-Zumino
case \cite{Fischler:2000fv}.

The presence of winding states can again be detected in the non-planar sector.
Indeed, by performing the integration over one set of momenta in eq.
(\ref{sym}),
one finds contributions to the free energy that are weighted by the length
of the
temporal circle. This can be seen for example by considering the following
factor
in eq. (\ref{sym}):

\begin{equation}
\int \frac{d^3 k}{(2\p)^3} \frac{e^{i\,p\, \t\, k}}
{\omega_k}  n_B(\omega_k) =   \sum_{n=1}^{\infty} \frac{1}{n^2\b^2 + (\t
p)^2}\;.
\end{equation}

We can again compare the limits $T^2\t \ll1$ and $T^2 \t \gg1$. In the case
where the
thermal wavelength is larger than the noncommutativity scale, the free energy
behavior is:

\begin{equation}
\frac{F}{V}  {\Big|_{\rm np}} \sim -g^2 N T^4.
\end{equation}
In the limit where the thermal wavelength is within the noncommmutative scale,

\begin{equation}
\frac{F}{V}  {\Big|_{\rm np}} \sim -g^2 N \frac{T^2 }{\t} \log
(T^2\t ).
\end{equation}
 This reveals again a reduction in the number of degrees of freedom
contributing to $F_{\rm np}$.

There seems to be a universal behavior
that whenever the inverse momenta fit into a Moyal cell, the associated
components of the field cannot be distinguished in their contributions to
$F_{\rm
np}$. What is then left over seems to be the
contribution due to a few quantum degrees of freedom per Moyal cell.

This picture will be further tested in the next section, where we will consider
a higher dimensional example.

\section{The thermodynamics of $ g\f^3 $ in $D=6$ dimensions}
Strictly speaking, this theory does not have good thermodynamic
behavior since it lacks a ground state. Therefore, in order to discuss the
thermodynamics of this system, we will take the coupling constant to be very
small and the temperature to be small enough\footnote{while still keeping
it much
larger than the mass} such that the excursions of the field
$\f$ away from the local minimum of the potential are within the bounded
region of the potential. This implies that the temperature satisfies the
inequality $T\ll{\frac{M}{g^{1/2}}}$.

The Lagrangian density for this system is:

$$ {\cal L} = \int {d^6x} (\,(\partial{\f})^2 - M^2 {\f}^2 - g\,\f*\f*\f ).$$

The free energy in this case has
no infrared divergences to leading order in an expansion in the coupling
constant g.
This is because the high dimensionality of the theory softens the IR
sensitivity of the free energy.

The non-planar contribution to the free energy at $O(g^2)$ is:

\begin{equation}
\frac{F}{V}  {\Big|_{\rm np}}= - g^2 T^2 \sum_{n,l}\int \frac{d^5
p}{(2\p)^5} \frac{d^5 k}{(2\p)^5}
\frac{e^{i\, p\, \t\, k}}
{(\frac{4\p^2 n^2}{{\b}^2} + k^2) (\frac{4\p^2 l^2}{{\b}^2} + p^2)
(\frac{4\p^2 (n
+ l)^2}{{\b}^2} + (k + p)^2)}\;.  \label{phi3raw}
\end{equation}

Again, one finds winding states in this theory. This can easily be seen by
rewriting eq. (\ref{phi3raw}) in the form\footnote{ for details, see
appendix \ref{appb}}:

\begin{equation}
\frac{F}{V}  {\Big|_{\rm np}}= - g^2 T \sum_{n,l}\int \frac{d^5 p}{(2\p)^5}
 \int_0^1 dx \int_{0}^{\infty} \frac{d{\alpha}_1}{{\alpha}_1^2} d{\alpha}_2 \,
e^{-(\alpha_2 +\alpha_1 x(1-x))(p^2 +
\frac{4\p^2 n^2}{\beta^2})} e^{-\frac{(\theta p)^2 + 4\p^2 l^2
\beta^2}{\alpha_1}} e^{2\p i n l x}\;,
\label{phi3}
\end{equation}
where we can think of $\alpha_2 +\alpha_1 x(1-x)$ as the proper time for the
propagator of momentum states and $\frac{1}{\alpha_1}$ as the proper time
for the
propogator of winding states.

Under the restriction that $\t_{0i} =0$, the most general case in $D=6$ is,
by a
convenient choice of coordinate system, given by nonvanishing $\t_{12}$
and
$\t_{34}$. If we take these two parameters to be of the same order of
magnitude,
one can distinguish two cases:
\begin{enumerate}
\item $T^2\ll  \frac{1}{\t_{12}},\frac{1}{\t_{34}} $
\item $T^2\gg  \frac{1}{\t_{12}},\frac{1}{\t_{34}}$.
\end{enumerate}
In the first case, the contribution to the free energy from the nonplanar
sector scales, as
a function of temperature, like
$T^6$. This is because the exponential involving the Moyal phase does not
oscillate much when
the momenta are distributed according to the thermal distributions.

In the high temperature limit where
$T^2\gg\frac{1}{\t_{12}},\frac{1}{\t_{34}}$, the possiblity of estimating the
behavior of the free energy using classical physics arises. Indeed, as
discussed above, if the system has many fewer degrees of freedom
contributing to
the $\t $ dependence of the free energy, as was the case in the four
dimensional
examples, then this approximation is valid.
Whether or not the approximation is valid is decided a posteriori: if the
remaining integrals are finite, then indeed classical statistical mechanics
is a
good approximation and gives the dominant contribution at high temperature.

Performing the classical statistical mechanics
calculation, i.e. evaluating the following expression,

\begin{equation}
\frac{F}{V}  {\Big|_{\rm np}} \sim  -g^2T^2 \int
\frac{d^5k}{(2\p)^5}\frac{d^5p}{(2\p)^5}\frac{e^{i\,p\,
\t\,k}}{p^2k^2(p+k)^2}\;,
\end{equation}
we find no UV divergences.

When $\t \sim\t_{12}\sim\t_{34}$,

\begin{equation}
\frac{F}{V}  {\Big|_{\rm np}} \sim  -g^2 \frac{T^2}{{{\t}_{34}}^2 -
{{\t}_{12}}^2}\log{\frac{\t_{12}}{\t_{34}}} \sim  -\frac{g^2T^2}{{\t}^2} \;.
\end{equation}

This behavior is again consistent with the picture on how degrees of freedom,
 as their
inverse momenta fall into Moyal cells, do not contribute to $F_{\rm np}$.

\section{Conclusions}

At high temperatures, one observes a substantial reduction in the number of
degrees of freedom that contribute to $F_{\rm np}$, the free energy due to the
non-planar sector of the theory. The picture that
emerges from the previous sections is
that this phenomenon can be traced to the modes of the fields with momenta
larger
than the
noncommutativity scales. What happens is that once the wavelengths are
within the Moyal
cells, there is no way to distinguish and count the separate contributions of
these modes to $F_{\rm np}$. What is left over seems to be the contribution of
a single degree of freedom per Moyal cell.

Because of this severe reduction in the number of degrees of freedom
contributing
to $F_{\rm np}$, this quantity can be calculated at high
temperature using classical statistical mechanics.

This behavior does not appear to depend on the details of the
noncommutative field
theory, the crucial element is the existence of a phase space structure for
space.

\acknowledgments
We thank Nathan Seiberg for very useful discussions.
The work of WF, EG, RMcN, AK-P, SP, PP is supported in part by the Robert
Welch Foundation and the NSF under grant number
PHY-9511632.
SP is also supported by NSF grant PHY-9973543.

\appendix

\section{The free energy of $N=4$, $D=4$ SYM} \label{appa}

We are working in Euclidean spacetime.

Let's begin with the commutative case.
We choose the decomposition of the 32 dimensional $\Gamma$-matrices given
in \cite{Osborn:1979tq}.
This yields the dimensionally reduced Lagrangian

\begin{eqnarray}
{\cal L} = &2 \,{\rm Tr}& \Big{(}-\frac{1}{4} F_{\mu \nu} F^{\mu \nu}
-\frac{i}{2}\bar{\chi}_K
\bar{\sigma}^\mu D_\mu \chi_K -\frac{1}{2} D_\mu \phi_i D^\mu \phi_i
-\frac{1}{2} D_\mu
\varphi_i D^\mu \varphi_i  \label{lsym} \\
& &- \frac{i}{2}g \alpha^i_{K L} ( \chi_K [ \phi_i, \chi_L] + \bar{\chi}_K
[ \phi_i,
\bar{\chi}_L] )  - \frac{i}{2}g \beta^i_{K L} ( -\chi_K [ \varphi_i,
\chi_L] + \bar{\chi}_K [
\varphi_i,\bar{\chi}_L] )   \nonumber  \\
& & -\frac{1}{4} g^2 ( [\phi_i,\phi_j][\phi_i,\phi_j] +
[\varphi_i,\varphi_j][\varphi_i,\varphi_j] +2
[\phi_i,\varphi_j][\phi_i,\varphi_j]) \Big{)}
\,,
\nonumber
\end{eqnarray}
where $F_{\mu \nu} = \partial_\mu A_\nu - \partial_\nu A_\mu -i g [A_\mu ,
A_\nu]$,
$D_\mu = \partial_\mu +ig[A_\mu,\cdot]$ and $\mu$, $\nu$ range from 1 to 4.
Remember that this theory contains one N=1 vector
multiplet and three N=1 chiral multiplets. Thus $i$ ranges from 1 to 3, $K$ and
$L$ from 1
to 4.
$\gamma_\mu$, $\alpha_i$ and $\beta_i$ satisfy
$\{\gamma_\mu,\gamma_\nu\}=-2 \delta_{\mu
\nu}$,
$\{\alpha_i,\alpha_j\}=-2 \delta_{i j}$, $\{\beta_i,\beta_j\}=-2 \delta_{i
j}$ and commute
among each other.

The diagrams that contribute to the free energy at two loop are depicted in
the following figure:

\epsfig{file=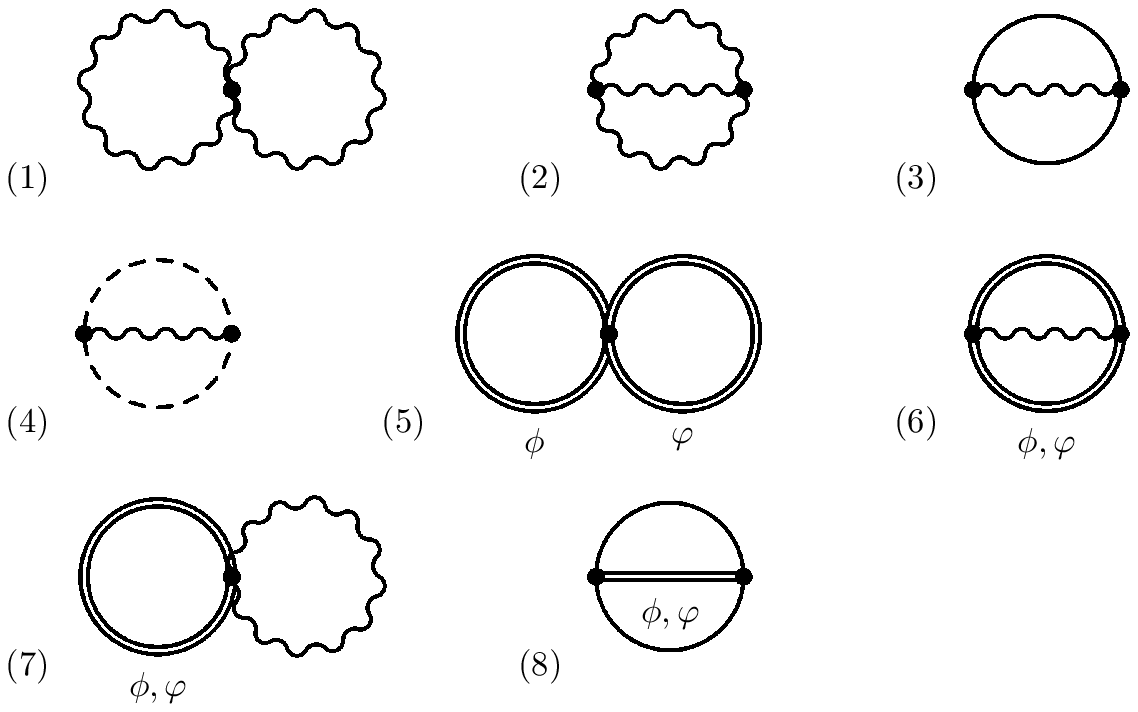}

At $T=0$, their contribution must vanish. It is easy to see that each diagram
is proportional
to $$g^2 f_{abc} f_{abc} \int \frac{d^4 k}{(2\pi)^4}\frac{d^4 p}{(2\pi)^4}
\frac{1}{k^2
p^2}.$$ The coefficients of the various contributions are given in the
following
table:

\begin{equation}
\begin{array}{|c|c|c|c|c|c|c|c|}
\hline
{} ~~1~~ & ~~2~~ & ~~3~~ & ~~4~~ & ~~5~~ & ~~6~~ & ~~7~~ & ~~8~~ \\
\hline
-3 & \frac{9}{4} & 4 & -\frac{1}{4} & \frac{9}{2} & \frac{9}{2} & -12
& 0 \\
\hline
\end{array}
\end{equation}
\\
We pass to finite
temperature by replacing the integration over energies by a sum over even or
odd Matsubara frequencies, for bosonic, fermionic degrees of freedom
respectively.

The noncommutative case does not require any additional calculation. We
obtain the
noncommutative theory by replacing the commutators in the Lagrangian
(\ref{lsym}) by Moyal
brackets.
This yields, for arbitrary fields $A$, $B$, which take values in the
fundamental representation, $$
[A,B]_*(x) =
\int
\frac{d^4 p}{(2\p)^4}
\frac{d^4 k}{(2\p)^4}
\tilde{A}^a(p)
\tilde{B}^b(k) e^{i(p+k)x} \left(  [t^a,t^b]\cos\frac{k\theta p}{2} +
i\{t^a,t^b\}\sin\frac{k\theta p}{2}\right).$$
For the gauge group $U(N)$, both the commutator and the anticommutator of
generators of the
fundamental close within the algebra: $$ [t^a,t^b]=if_{abc}t^c \:,
\:\{t^a,t^b\}=d_{abc}t^c.$$ The noncommutative result is thus obtained from
the commutative
one by replacing~\cite{Arcioni:2000hw} $$f_{abc}  \rightarrow f_{abc}
\cos\frac{k\theta
p}{2} +
 d_{abc} \sin\frac{k\theta p}{2} \,. $$
To obtain (\ref{sym}), we need to perform the sums $f_{abc} f_{abc}$ and
$d_{abc}d_{abc}$.
The first is of course identical to the $SU(N)$ result, $N(N^2-1)$. The
second can be
obtained from the SU(N) result, $N(N-\frac{4}{N})$, where
$\{t^a,t^b\}=d^{SU(N)}_{abc}t^c
+\frac{1}{N}
\delta_{ab}$, by including the $U(1)$ generator
$t^{N^2}=\frac{1}{\sqrt{2N}}1$. This gives
$N(N^2+1)$.

\section{The free energy of the $D=6$, $g\f^3$ theory}  \label{appb}
We will first briefly show how the winding states appear in this case, then
we will
perform the classical statistical mechanics calculation of the non-planar
contribution to
$O(g^2)$ to the free energy.

The nonplanar contribution to the free energy to
$O(g^2)$ is

\begin{equation}
-g^2T^2\sum_{n,l}\int \frac{d^5p}{(2\pi)^5}\,
\frac{d^5k}{(2\pi)^5}\
\frac{e^{i(\theta_{12}
(p_1\, k_2-p_2\, k_1)+\theta_{34}(p_3\, k_4-p_4\, k_3))}}
{(p^2+\frac{4\pi^2 n^2}{\beta^2})
(k^2+\frac{4\pi^2 l^2}{\beta^2})( (p+k)^2+\frac{4\pi^2 (n+l)^2}{\beta^2})}.
\label{sixdimensions}
\end{equation}

The appearance of winding contributions in this integral can be shown by
introducing a
Feynman parameter, $x$, and Schwinger parameters ${\alpha}_1$ and ${\alpha}_2$:

\begin{eqnarray}
-g^2T^2\sum_{n,l}\int \frac{d^5p}{(2\pi)^5}\,
\frac{d^5k}{(2\pi)^5} & & \int_0^1 dx \int_0^\infty d \alpha_1 d \alpha_2 \,
\alpha_1  \,  \\
& & e^{-\alpha_2 (p^2+\frac{4\p^2 n^2}{\beta^2})}  e^{-\alpha_1 k^2 - \alpha_1
(p^2 x+2 p k x + 4\p^2
  \frac{n^2 x +2nlx + l^2}{\beta^2} )} e^{i\,p \, \t \, k}. \nonumber
\end{eqnarray}
Performing the integral over $k$ followed by a Poisson resummation over $l$
gives

\begin{equation}
\frac{F}{V}  {\Big|_{\rm np}}= - g^2 T \sum_{n,l}\int \frac{d^5 p}{(2\p)^5}
 \int_0^1 dx \int_{0}^\infty \frac{d{\alpha}_1}{{\alpha}_1^2} d{\alpha}_2 \,
e^{-(\alpha_2 +\alpha_1 x(1-x))(p^2 +
\frac{4\p^2 n^2}{\beta^2})} e^{-\frac{(\theta p)^2 + 4\p^2 l^2
\beta^2}{\alpha_1}} e^{2\p i n l x}\;,
\end{equation}
which is equation (\ref{phi3}).

We now turn to evaluating the dominant contribution to this expression at
high temperature.
As explained in the text, this is given by the
$n=l=0$ contribution in equation (\ref{sixdimensions}). We will for
simplicity limit our
discussion to this mode. We introduce Feynman and Schwinger parameters:
\begin{eqnarray}
-g^2T^2\int_0^\infty d\alpha \, \alpha^2\int^1_0 dx
\int^{1-x}_0 dy\int d^5p\, d^5k\,
e^{-\alpha (
(1-y)p^2+ (1-x) k^2 + 2 (1-x-y)p . k) + i p\,\theta\, k}.
\label{parameters}
\end{eqnarray}
The Gaussian integrals can now be performed with the result:
\begin{eqnarray}
-g^2 T^2\int_0^\infty\alpha\,d\alpha\int^1_0 dx \int^{1-x}_0 dy\,
\frac{1}{f(x,y)^{1/2}}\, \frac{1}{\alpha^2 f(x,y) +
\theta_{12}^2}\, \frac{1}{\alpha^2 f(x,y) + \theta_{34}^2},
\label{gaussian}
\end{eqnarray}
where $f(x,y) = x (1-x) + y (1-y) -x \, y$.
The Feynman and Schwinger integrals can also be done exactly and they give:
\begin{equation}
-g^2 T^2\, \frac{1}{\theta_{12}^2 -
\theta_{34}^2}\, \log\left(\theta_{12}/\theta_{34}\right).
\label{result}
\end{equation}
This is the general result when $M/g^{1/2}\gg T\gg M$.

We note in passing that the zero temperature
vacuum energy of this theory is finite:
\begin{equation}
g^2\int \frac{d^6p}{(2\pi)^6}\,\frac{d^6k}{(2\pi)^6}\
\frac{e^{-i(\theta_{12}
(p_1\, k_2-p_2\, k_1)+\theta_{34}
(p_3\, k_4-p_4\, k_3))}}{p^2\, k^2\,( p+k)^2} =
g^2 \frac{1}{\theta_{12}+\theta_{34}}\, \frac{1}{\theta_{12}\, \theta_{34}}.
\label{vacuum}
\end{equation}

\end{document}